\documentclass[pre,superscriptaddress,twocolumn,a4paper,showpacs]{revtex4-1}
\usepackage{graphics}
\usepackage{amssymb}
\usepackage{wasysym}
\usepackage{latexsym}
\usepackage{graphicx}
\usepackage{epsfig}
\usepackage{subfigure}
\begin{document}

\title{The surprising little effectiveness of cooperative algorithms in parallel problem solving}

\author{Sandro M. Reia}
\affiliation{Instituto de F\'{\i}sica de S\~ao Carlos,
  Universidade de S\~ao Paulo,
  Caixa Postal 369, 13560-970 S\~ao Carlos, S\~ao Paulo, Brazil}

\author{Larissa F. Aquino}  
\affiliation{Instituto de F\'{\i}sica de S\~ao Carlos,
  Universidade de S\~ao Paulo,
  Caixa Postal 369, 13560-970 S\~ao Carlos, S\~ao Paulo, Brazil} 

\author{Jos\'e F.  Fontanari}
\affiliation{Instituto de F\'{\i}sica de S\~ao Carlos,
  Universidade de S\~ao Paulo,
  Caixa Postal 369, 13560-970 S\~ao Carlos, S\~ao Paulo, Brazil}

\begin{abstract}
Biological and cultural inspired optimization algorithms are  nowadays part of the basic toolkit of  a great  many research domains. 
By mimicking  processes in nature and animal societies,  
these general-purpose search algorithms  promise to deliver optimal or near-optimal solutions using hardly any  information on the optimization problems they are set to tackle. Here  we study the performances of a cultural-inspired algorithm -- the imitative learning search -- as well as of asexual and sexual variants of evolutionary algorithms in finding the global maxima of  NK-fitness landscapes.
The main  performance measure is the total number of agent updates required by the algorithms  to find those global maxima  and  the baseline performance, which  establishes the effectiveness of the cooperative algorithms,  is set by  the blind search in which the agents explore the problem space (binary strings) 
 by  flipping bits at random. We find that even for smooth landscapes that exhibit a single maximum, the evolutionary algorithms  do not perform  much better than the blind search due to the stochastic effects of the genetic roulette. The imitative learning is immune to this effect thanks to the deterministic choice of the fittest string in the population, which is used as a model for imitation. The tradeoff is that  for rugged landscapes the imitative learning search  is more prone to be trapped in local maxima than the evolutionary algorithms. In fact, 
in the case of  rugged landscapes with a mild density of local maxima,  the blind search either beats or matches  the  cooperative algorithms regardless of whether the task is to find the global maximum or to find the fittest state within a given runtime.
\end{abstract}

\maketitle

\section{Introduction}\label{sec:intro}

Today's   web-enabled collective intelligence enterprises such as Google and Wikipedia \cite{Malone_10}  are outstanding implementations of  the familiar notion that  the solution  of important real-world problems is beyond the capability of any  single individual and requires  the cooperative  effort  of  many individuals.  In fact, the benefits of cooperative work  to tackle problems that endanger survival  have long been explored by nature \cite{Queller_09} and nature's diverse strategies  have, in turn, been developed into a variety of general-purpose  optimization algorithms \cite{Kar_16,Gao_19}. 

Perhaps, the first and most popular of these bio-in\-spired algorithms are the evolutionary algorithms \cite{Holland_92} that rely on the well-known biological processes of mutation, selection and recombination to drive a population  towards global or near-global maxima of abstract fitness landscapes. In this line, there are also the more recent and less known cultural-inspired algorithms  \cite{Kennedy_98,Fontanari_14}, where   social learning  (imitation)  replaces   the  biological processes of selection and recombination \cite{Blackmore_00,Boyd_05}.

Both  the evolutionary and the cultural algorithms are examples of  parallel or distributed cooperative problem-solving systems \cite{Huberman_90,Clearwater_91} in which a number of equivalent agents seek to solve the same problem  and  the activities of a particular agent offer insight to others about the configuration of the problem space \cite{Lazer_07}. This is typically achieved through the exchange of information among the agents about  their partial success (i.e., their states and their fitness at the current trial)  towards the completion of the task. 

Here we study the performances of the cultural-inspired imitative learning search \cite{Fontanari_14,Fontanari_15} and of   evolutionary algorithms \cite{Goldberg_89,Back_96}  for the  problem of  finding the global maxima of  NK-fitness landscapes \cite{Kauffman_89}.  
The main advantage of using this  problem is the possibility of tuning  the ruggedness of the landscape, which is roughly determined  by the number of local maxima. In addition, the implementations of  the cultural and evolutionary algorithms  to explore the landscape are straightforward  since the state space of  the NK-fitness landscapes are binary strings of length $N$. (Hence, in this paper  we  will use  the terms agent and string interchangeably.)  We  take the  blind search, in which the agents simply flip bits at random independently of each other, as a baseline  to gauge the efficiency of the cooperative search algorithms. Although our main measure of performance is the number of agent updates required to find the global maxima, which  is essentially the computational cost of the search,  we considered also other measures such as the probability of finding the global maximum and the fittest string found   for a fixed runtime. 
  
The cooperative search algorithms we consider here have only two tunable parameters, viz., the population size $M$ and the bitwise mutation probability $u$. In particular, in the imitative learning (IL) search the agents imitate  the model agent -- the fittest agent in the population at the generation -- by  copying one of its bits. The resulting string then goes through the mutation process where each bit  is flipped   with probability $u$.  We consider two variants of the evolutionary algorithms, namely, the asexual variant (AGA) that accounts for  mutation and selection,  and the sexual variant (SGA) that accounts for recombination  as well.  The blind search corresponds to the choice $u=1/2$ in any of these algorithms.

Surprisingly, we find that for simple problems in which the fitness landscapes are smooth and exhibit a single maximum, the evolutionary algorithms do not perform  much better than the blind search. This is probably because the genetic roulette is not effective to select the fittest agent in the case the agents have similar fitness values. The genetic drift  effect  becomes stronger as the population size decreases and for small sizes the evolutionary algorithms typically perform worse than the blind search. This finding exposes the pitfall of  general-purpose optimization  algorithms that  use little or no information regarding   the  optimization problem they are set to solve. 

The IL search is immune to genetic drift because the model agent is always chosen as the fittest string in the population. The tradeoff is that IL is strongly affected by the trapping  effects of the local maxima in rugged landscapes, so it performs much worse than the blind search in the case of  low mutation probability or  large population size.
Nevertheless, tuning $M$ and $u$ independently for the three cooperative search algorithms indicates that IL is either superior or equivalent to   the evolutionary algorithms  for rugged landscapes.  In addition,  we find that already for mildly rugged landscapes, the blind search either outperforms or matches the cooperative algorithms.  These conclusions holds true even when  the task of the search algorithms is to find the fittest state within a relatively short runtime so that the chances of reaching the global maximum are negligible.

The rest of this paper is organized as follows. Section \ref{sec:NK}  presents the NK model of  rugged fitness landscapes and Section \ref{sec:PSA} describes the three cooperative search algorithms, as well as  the blind search, and introduces our definition of computational cost. Section \ref{sec:Res} offers a comparison and  discussion of  the performances of the search algorithms on smooth and rugged fitness landscapes. Section \ref{sec:Conc} summarizes our main findings and offers our concluding remarks.

\section{NK-fitness landscapes}\label{sec:NK}

The NK model \cite{Kauffman_89} is the choice computational implementation  of fitness landscapes that 
has been extensively used to study optimization problems in theoretical immunology, population genetics, developmental biology and protein folding \cite{Kauffman_95}. Although the NK model was  widely used   to study  adaptive evolution  as walks on rugged fitness landscapes,   its repute went  way beyond the (theoretical) biology realm. In fact,  today the NK model is considered
a paradigm  for problem representation in management research \cite{Lazer_07,Levinthal_97,Billinger_13},
 as it allows the manipulation of the  difficulty of the problems and challenges posed to  individuals and companies.
 
 More pointedly,
the NK model is defined in the space of binary strings of length $N$ and so the parameter $N$ determines the size of the state space, $2^N$.  
The other parameter  $K =0, \ldots, N-1$  determines the range of the epistatic interactions among the bits of the binary string and  influences strongly the number of local maxima on the landscape. We recall that   two bits are said to be epistatic whenever the combined effects of their contributions to the fitness of  the binary string  are not  merely additive. In particular,
for $K=0$ the smooth and additive landscape has one single maximum whereas for $K=N-1$, the (uncorrelated) landscape  has on the average  $2^N/\left ( N + 1 \right)$ maxima with respect to single bit flips \cite{Kauffman_87}. Since the $2^N$ binary strings can be arranged in a $N$-dimensional hypercube, we can say that $N$ is the dimensionality of the landscape.

In the NK model, each string  $\mathbf{x} = \left ( x_1, x_2, \ldots,x_N \right )$ with
$x_i = 0,1$ has  a fitness value $\Phi \left ( \mathbf{x}  \right ) $ that is given by  the average  of the contributions  of each 
component $i$ in the string, i.e.,
\begin{equation}\label{eq:Phi}
\Phi \left ( \mathbf{x}  \right ) = \frac{1}{N} \sum_{i=1}^N \phi_i \left (  \mathbf{x}  \right ) ,
\end{equation}
where $ \phi_i$ is the contribution of component $i$ to the  fitness of string $ \mathbf{x}$. It is assumed that $ \phi_i$ depends on the state $x_i$  as well as on the states of the $K$ right neighbors of $i$, i.e., $\phi_i = \phi_i \left ( x_i, x_{i+1}, \ldots, x_{i+K} \right )$ with the arithmetic in the subscripts done modulo $N$.  Hence $K$ measures the degree of interaction (epistasis) among the components of the bit string.  Here we assume, in addition, that  the functions $\phi_i$ with $i=1, \ldots, N$ are distinct real-valued functions on $\left \{ 0,1 \right \}^{K+1}$ and, as usual,  we assign to each $ \phi_i$ a uniformly distributed random number  in the unit interval \cite{Kauffman_89}. Because of the randomness of $\phi_i$, we can guarantee that  $\Phi   \in \left ( 0, 1 \right )$ has a unique global maximum and that different strings have different fitness values. 

There are many variants of the NK-model characterized by different interaction structures, i.e., different ways of choosing the $K$ interaction partners of a site (see \cite{Hwang_18} for a recent review). Here we consider the adjacent neighborhood variant only,  which is one  of Kauffman's original choices.  Whereas in many cases different interaction structures give rise to similar behavior, they affect the computational complexity, and hence the efficiency of searches, in a nontrivial way \cite{Wright_00}. In particular, for the adjacent neighborhood variant there is a dynamic programming algorithm that is polynomial in  $N$  and exponential in $K$.  For  the random version, where the $K$ interaction partners are  chosen randomly, the optimization of the NK-fitness landscapes for $K \geq 2$  is NP-complete \cite{Wright_00}, which  means that the time required to solve some particular instances of the  problem using any currently known deterministic algorithm increases exponentially fast with the length $N$ of the strings \cite{Garey_79}.

As pointed out, for $K=0$ there are no local maxima and the sole maximum of $\Phi$ is easily located by picking for each component $i$ the state $x_i = 0$ if  $\phi_i \left ( 0 \right ) >  \phi_i \left ( 1 \right )$ or the state  $x_i = 1$, otherwise. 
Increase of the parameter $K$ from $0$ to $N-1$  results in the decrease of the correlation between the fitness of neighboring strings 
(i.e., strings that differ at a single component) in the state space.
For $K=N-1$, those fitness values are  uncorrelated so the NK model reduces to the house-of-cards landscape \cite{Kauffman_87,Macken_89}.  The simplest way to see this is to consider  two neighboring configurations, say
$\mathbf{x}^a = \left ( 0, 0, \ldots, 0 \right )$ and $\mathbf{x}^b = \left ( 1, 0, \ldots, 0 \right )$, and  calculate explicitly the correlation between their fitness. This procedure yields 
\begin{equation}\label{eq:corr}
  \text{corr} \left ( \Phi \left ( \mathbf{x}^a  \right ),   \Phi \left ( \mathbf{x}^b  \right ) \right ) = 1 - \frac{K+1}{N},
\end{equation}
indicating thus that the increase of the dimensionality of the landscape $N$ while the epistasis parameter $K$ is kept fixed  produces nearly flat fitness landscapes.

We note that since the functions  $ \phi_i$ are random, the ruggedness measures (e.g., the number of local maxima)  of a particular realization of a NK landscape is not uniquely determined by the parameters $N$ and $K$. In fact, the number of local maxima  can vary considerably between landscapes characterized by the same values of $N$ and $K>0$ \cite{Kauffman_89}, which implies that the  performance of any search  algorithm based on the  local correlations of the fitness landscape will depend on the particular realization of the landscape. Therefore,  in order to  produce a  meaningful  comparison between the search algorithms we must guarantee that they survey the same landscapes. To  achieve that  we generate and store a set of 100 landscape realizations for each value of $N$ and $K$, which are then  used to test the parallel search algorithms.

\section{Parallel search algorithms}\label{sec:PSA}

Here we describe the three cooperative parallel search algorithms we use to explore the  NK-fitness landscapes, namely, the imitative learning (IL) search,  the asexual genetic algorithm (AGA) and the sexual genetic algorithm (SGA).  In order to render possible  a fair comparison between these algorithms we implement   a slight variant of the IL: instead of considering mutation and imitation as two independent processes as in the original version  \cite{Fontanari_14,Fontanari_15}, here we tie these two processes together so that mutation takes place after imitation, thus mimicking the usual evolutionary view of mutations as copy errors.
For the sake of completeness, we present also the blind independent search where the agents explore the fitness landscape  flipping bits at random.

We consider a well-mixed population of $M$ agents or binary strings of length $N$ that explore the state space of an  NK-fitness landscape searching for its unique  global maximum. Initially, all binary strings are  drawn at  random with equal probability for the digits $0$ and $1$. The $M$ agents are updated synchronously following the rules of the specific search algorithm and the search is halted when one of the agents finds the global maximum. We denote by $t^*$ the time when this happens.

\subsection{Imitative learning (IL)} 

In the imitative learning search, the synchronous update of the $M$ agents proceeds as follows.
At time $t$ we  first determine the model agent (i.e., the fittest agent  in the population) and then we repeat the following update rule $M$ times before incrementing the time to $t+1$. 

The update rule consists of selecting a  string at random with uniform probability (the target string) which will then imitate the model string.  
More pointedly, the  model string and the target string   are compared  and the different bits are singled out.  Then one  of the distinct bits in the target string is selected at random and flipped so that this bit is now the same in both strings.  This imitation procedure is inspired by
the mechanism used to model the influence of external media \cite{Shibanai _01,Avella_10,Peres_11} in  the celebrated agent-based model proposed by  Axelrod to study the process of culture dissemination \cite{Axelrod_97}. After imitation the target string goes through the mutation process: each  of its $N$ bits is flipped with probability $u$ so that the mean number of flipped bits is $Nu$. This is the usual mutation operation of the evolutionary algorithms. The resulting string is then passed to the next generation.

As expected, imitation  results in  the increase of the similarity between the target and the model strings, which may not necessarily lead to  an increase of the fitness of the target string.   If the target string  is identical to the model string,  which is not an  uncommon situation since the imitation process reduces the diversity of the population, then  imitation does not occur and the target string changes due to mutation only. This means that for $u=0$ the population rapidly becomes isogenic and the search for the global maximum is very likely to  fail.
We note  that  a same string may be chosen as target string more than once and,  more importantly, that there is no guarantee that the  model string will pass to the new generation, as opposed to the elitist selection strategy of some evolutionary algorithms \cite{Back_96}.

\subsection{Asexual genetic algorithm (AGA)} 

In the context of evolutionary algorithms, the parallel update of the agents amounts to the usual assumption of non-overlapping generations in which the offspring replace the parental population in a single time step.  Accordingly, at time $t$  we repeat the following  update procedure $M$ times and then  increment the time  from $t$ to $t+1$.  
As usual in evolutionary algorithms, the  two operations --  replication  and mutation --  are applied sequentially \cite{Back_96}. Explicitly, we  select a string with probability proportional to its fitness and then subject it to the mutation operation described before. The resulting string is then passed  to the next generation.
 We note that  the same string may be chosen more than once and that new strings are produced by the mutation operation only. The inefficiency of AGA to generate and maintain string diversity  is  the culprit for its  poor performance  reported in this paper.

\subsection{Sexual genetic algorithm (SGA)} 

In this case the two operations that  comprise the update rule are reproduction and mutation. Reproduction consists in selecting two  strings (parents)  without replacement and with probability proportional to their fitness.
 The single offspring of each mating is generated by applying the one point crossover operation: we pick one point $ 1 \leq n \leq N-1$  at random from each of parents' strings to form one offspring string by taking all bits from the first parent up to the crossover point $n$, and all bits from the second parent beyond the crossover point.  Thus  the offspring will always be a recombinant string. Next, the offspring is  put through the mutation process as before and then passed to the next generation. This update procedure is repeated $M$ times before incrementing the time variable by one unit. As in the previous algorithms, the  same string can be selected more than once as a parent in the reproduction process. In the SGA, string diversity is generated both by the mutation and the crossover operations.
 
We use the  word reproduction rather than replication in the SGA  because the offspring may be different from their parents (it is equal in the case the parents are clones) in contrast to the AGA where, except for the bits flipped in the mutation operation,  the offspring is always a clone of the parent.
 
\subsection{Blind search (BS)}\label{BS}

In the blind search the agents flip bits at random (i.e., $u = 1/2$) and so the ruggedness of the landscape has no effect whatsoever on their chances to find the global maximum, which depends only on the length of the strings $N$ and on the population size $M$. Since the agents explore  the fitness landscape independently of each other, the halting time  of the search is given by 
\begin{equation}\label{TM}
t^* = \min \left ( t_1^*, \ldots, t_M^* \right ),
\end{equation}
where $t_i^*$, $i=1, \ldots, M$,  are  identically distributed  independent random variables distributed by the geometric distribution
\begin{equation}
f \left ( t_i^* \right ) =  p \left ( 1- p \right)^{t_i^*-1}.
\end{equation}
Here  $p= 1/2^N$ is the success probability. (We recall that a NK-fitness landscape has a unique global maximum.) The probability distribution of the halting time $t^*$ is also a geometric distribution
\cite{Feller_68} with success probability $1 - \left ( 1 - p \right)^M$, i.e.,
\begin{eqnarray} \label{eq:PM}
P_M \left ( t^*  \right ) & = & \left [ 1 - \left ( 1 - p \right)^M \right ] \left ( 1 - p \right)^{M \left ( t^* - 1 \right )}    \\
& \approx & M p \exp \left ( - Mp t^* \right )  ,
\end{eqnarray}
where in the last step we have assumed that the system size is much smaller than the size of the state space, i.e., $Mp \ll 1$.  
The mean time to find the global maximum is then
\begin{eqnarray}
\langle t^*  \rangle & = &  \frac{1}{  1 - \left ( 1 - p \right)^M } \label{eq:t} \\
& \approx & \frac{1}{Mp} \nonumber ,
\end{eqnarray}
where, as before, the last step assumes that  $Mp \ll 1$.
It is also of interest to know the probability that the population of $M$ agents finds the global maximum before or at time $t$, which is given by
\begin{equation}\label{eq:PiM}
\pi_M \left (t \right ) =  \sum_{t^*=1}^{t} P_M \left (t^* \right )  =  1 - \left ( 1 - p \right)^{Mt}  .
\end{equation}

\subsection{Computational cost}

We measure the computational cost of a search by the total number of string updates  performed by the algorithms until they  find the global maximum of the NK-fitness landscape. Hence we ignore the complexity of the update procedures which may greatly impact the 
actual running time of the algorithms in a computer. Needless  to say, this performance measure is very unfavorable to the blind search, which has the simplest and fastest update rule. Of course, the total number of updates is simply $M t^*$ where $ t^*$ is the halting time of the algorithm. Since $t^*$ is typically on the order of $ 2^N$ and in order to compare performances for landscapes of different dimensionality,  we choose to define the computational cost  $C$ of a search as
\begin{equation}\label{C}
 C  = M t^*/  2^N  .
\end{equation}
This quantity must be averaged over many searches in a same landscape and the result then averaged over an ensemble of landscapes with the same  parameters $N$ and $K$. 

For the blind search, the mean computational cost is simply 
\begin{equation}\label{Cind}
\langle C \rangle = \frac{Mp}{ \left [ 1 - \left ( 1-p \right)^M \right ]} ,
\end{equation}
where    the  notation $\langle \ldots \rangle$  stands for the average over independent searches on the same landscape. Since all landscapes with fixed $N$ and a single global maximum are equivalent from the perspective of the blind search, there is no need to  average over different landscapes in equation (\ref{Cind}).
In particular, for $Mp \ll 1$ we have 
$\langle C \rangle \approx  1 $ and for $Mp \gg 1$ we have
$\langle C \rangle \approx M/2^N$. The first and more realistic regime is  characterized by a  mean computational cost that is independent of the population size $M$ and  corresponds to the case that  the halting time $t^*$ decreases linearly with increasing $M$. The second regime, where $\langle C \rangle$  increases linearly with $M$, corresponds to the situation  $t^* \approx 1$, i.e., the population size is so large that the global maximum is likely to be found already during the assemblage of  the initial  population.

\begin{figure*}[t!]
\centering  
\includegraphics[width=0.85\textwidth]{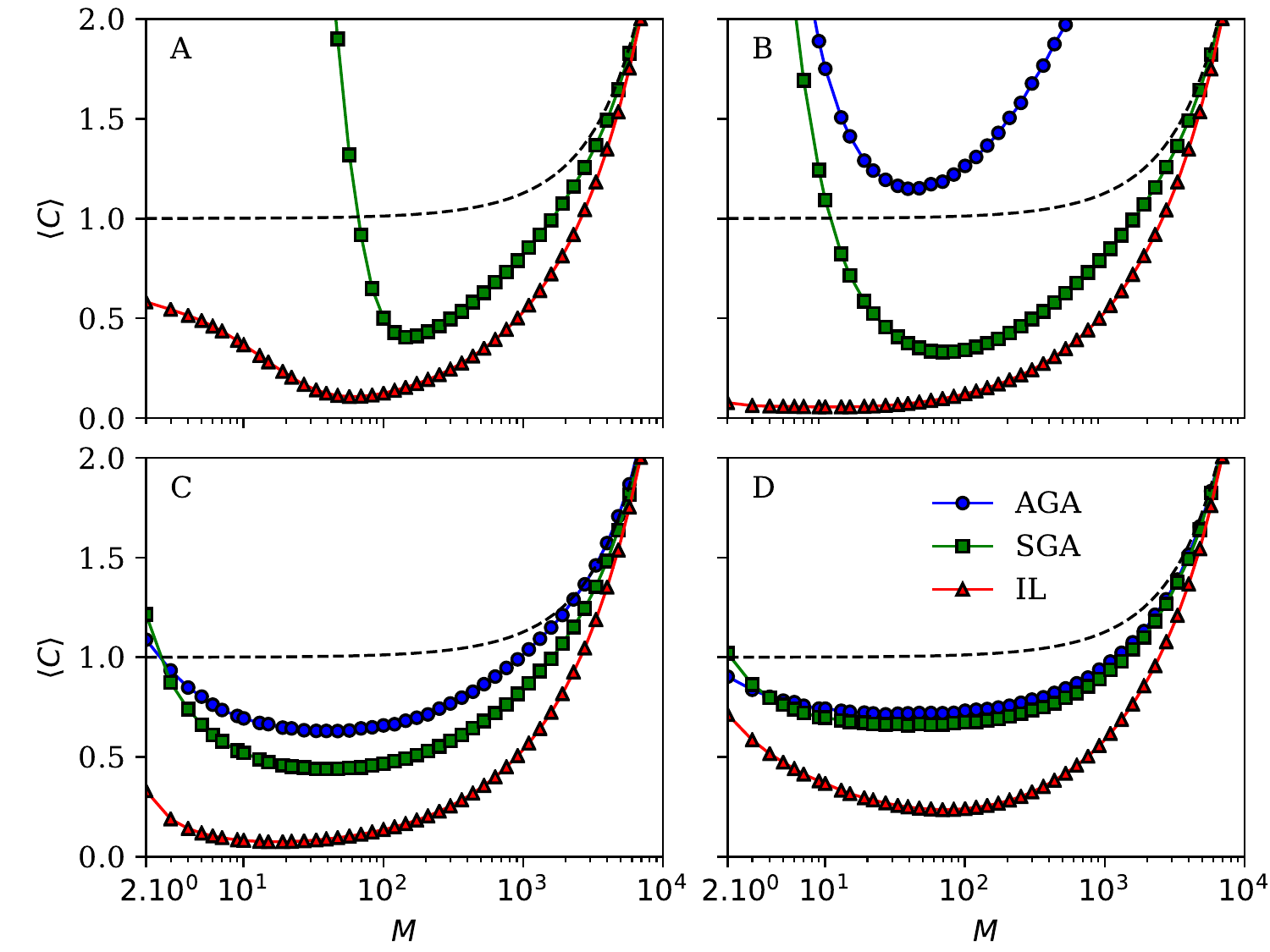}
\caption{Mean computational cost $\langle C \rangle$ as function of the  population size $M$ for smooth ($K=0$)  landscapes with dimensionality $N=12$. The bitwise mutation probability is $u=0.001$ (panel A),  $u=0.01$ (panel B),  $u=0.1$ (panel C), and $u=0.2$ (panel D). 
  The dashed curve is the analytical prediction for the blind search, equation (\ref{Cind}), and the symbols are the simulation results for the imitative learning (IL) search, the asexual (AGA) and the  sexual (SGA)  genetic algorithms as indicated. }
\label{fig1}
\end{figure*}

\section{Results}\label{sec:Res}

As pointed out, the performances of the cooperative search algorithms are  measured by the mean  computational cost $\langle C \rangle$, which  is estimated by averaging the computational cost defined  in equation (\ref{C}) over $10^3$  searches on the same  landscape realization.  The resulting average cost is then further averaged over the set of 100 landscape realizations with the same values of the parameters $N$ and $K$. Whereas the results for the cooperative search algorithms are obtained  via simulations, the results for the blind search are given by the exact analytical  expressions derived before, unless  otherwise stated.

\begin{figure*}[t!]
\centering  
\includegraphics[width=0.85\textwidth]{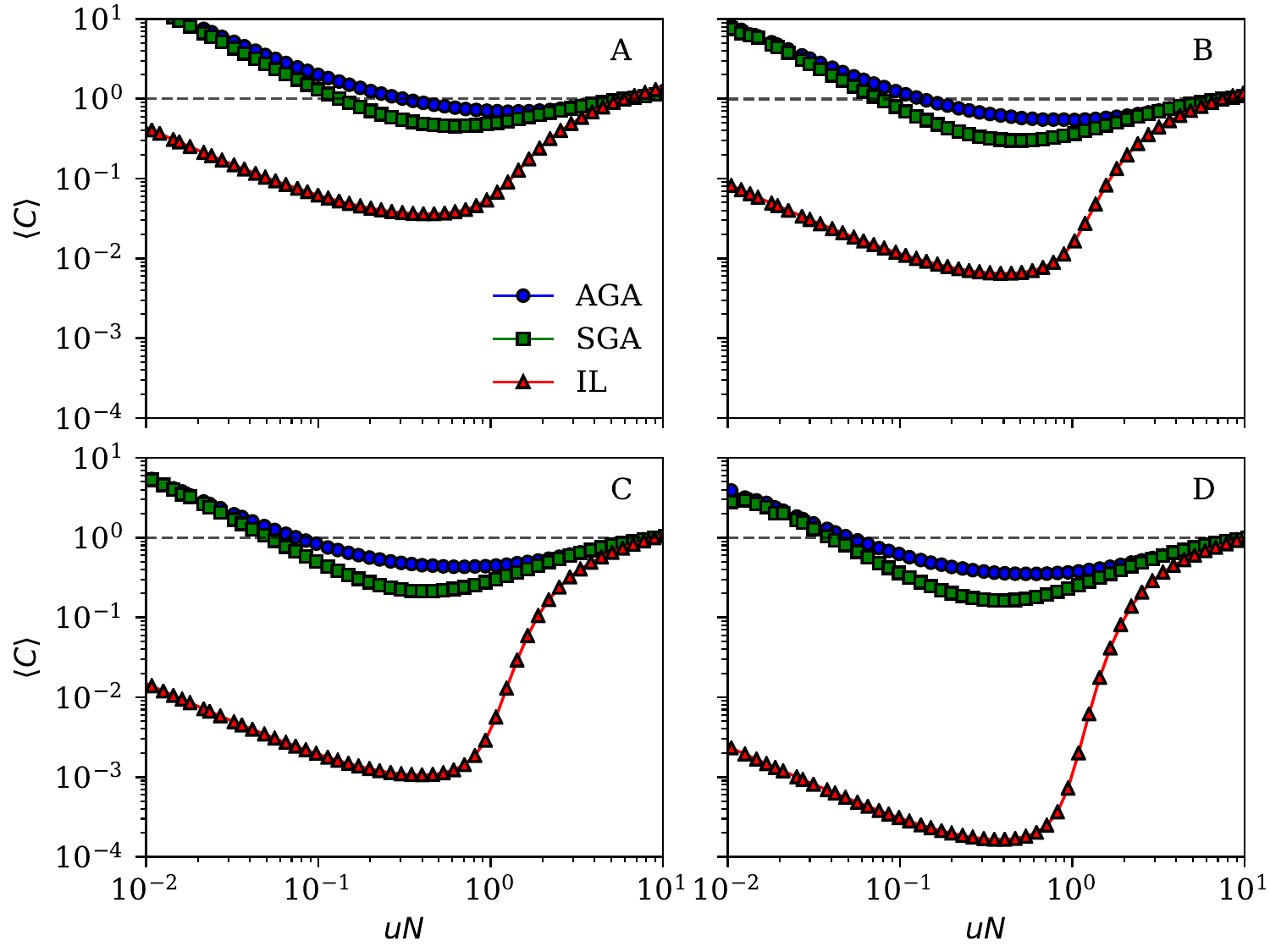}
\caption{Mean computational cost $\langle C \rangle$ as function of the mean number of mutations per string $u N$ for smooth ($K=0$) landscapes with dimensionality  $N=12$ (panel A),  $N=15$ (panel B), $N=18$ (panel C) and    $N=21$ (panel D).
 The  population size is $M=10$.  The dashed straight lines indicate the performance of the blind search and the symbols are the simulation results for the imitative learning (IL) search, the asexual (AGA)  and the  sexual  (SGA) genetic algorithms as indicated.   }
\label{fig2}
\end{figure*}

\subsection{Smooth landscapes}

Regardless of the dimensionality $N$, the  NK-fitness landscape with $K=0$  exhibits a single maximum. Hence, in principle,  finding this maximum should be a very easy task to powerful search algorithms such as AGA and SGA. Somewhat surprisingly, this is not the case, as we will see next.

  Figure \ref{fig1} summarizes the performances of the three cooperative search algorithms, viz. IL search,  AGA and  SGA,  as well as of the blind search, for smooth landscapes with fixed dimensionality $N=12$. The performance of the blind search is used as 
 a baseline to determine the usefulness of the cooperative algorithms. This figure reveals a few surprising results. Although the poor performance of the evolutionary algorithms for small population sizes was somewhat expected since the genetic drift (i.e., the stochastic effects due to the finitude of the population) overwhelms the selective pressure towards fitter strings, it comes as a surprise that those algorithms perform  much worse than the blind search for small $u$. (The AGA does not appear in panel A of  Figure \ref{fig1} because its mean computational cost is greater than 2 for all $M$.) The IL  search does not suffer from the drift effect since it always picks the fittest string to imitate, i.e., this choice is not probabilistic as in the evolutionary algorithms. 
 All the cooperative search algorithms considered exhibit an optimum population size that minimizes the computational cost of the search. This is due to the duplication of work (i.e., the presence of multiple copies of a same string) that occurs for large $M$ and reduces the efficiency of the search.

 We note that for low dimensional landscapes the performances of the three cooperative search algorithms are not remarkably better than the baseline set by the blind search. More explicitly, by fine tuning  the parameters $M$ and $u$, the IL  search yields a computational cost that is  about 20 times lower than  the baseline, whereas the SGA  and AGA  result  in a twofold improvement   over the baseline only.   For purpose of comparison,  we  note that a greedy heuristics starting from a random string  yields $\langle C \rangle = N/2^{N+1} \approx 0.0015$ for $N=12$ that is about 750 times lower than the baseline.

 Figure \ref{fig2}  shows that the performance of the evolutionary algorithms scales very poorly with increasing dimensionality $N$, whereas the computational cost of the IL search
decreases exponentially with increasing $N$, similarly to the greedy heuristics. In addition, this figure reveals that for each cooperative search algorithm there is a mean number of mutations per string  $uN$ that minimizes the computational cost for fixed population size $M$. This is expected since for  $uN = N/2$, the performances of the cooperative search algorithms  are identical to the baseline by construction, whereas for $uN \to 0$ they  may  not find the solution for some initial population settings, thus leading to the divergence of the mean computational cost.

\begin{figure*}[t!]
\centering  
\includegraphics[width=0.88\textwidth]{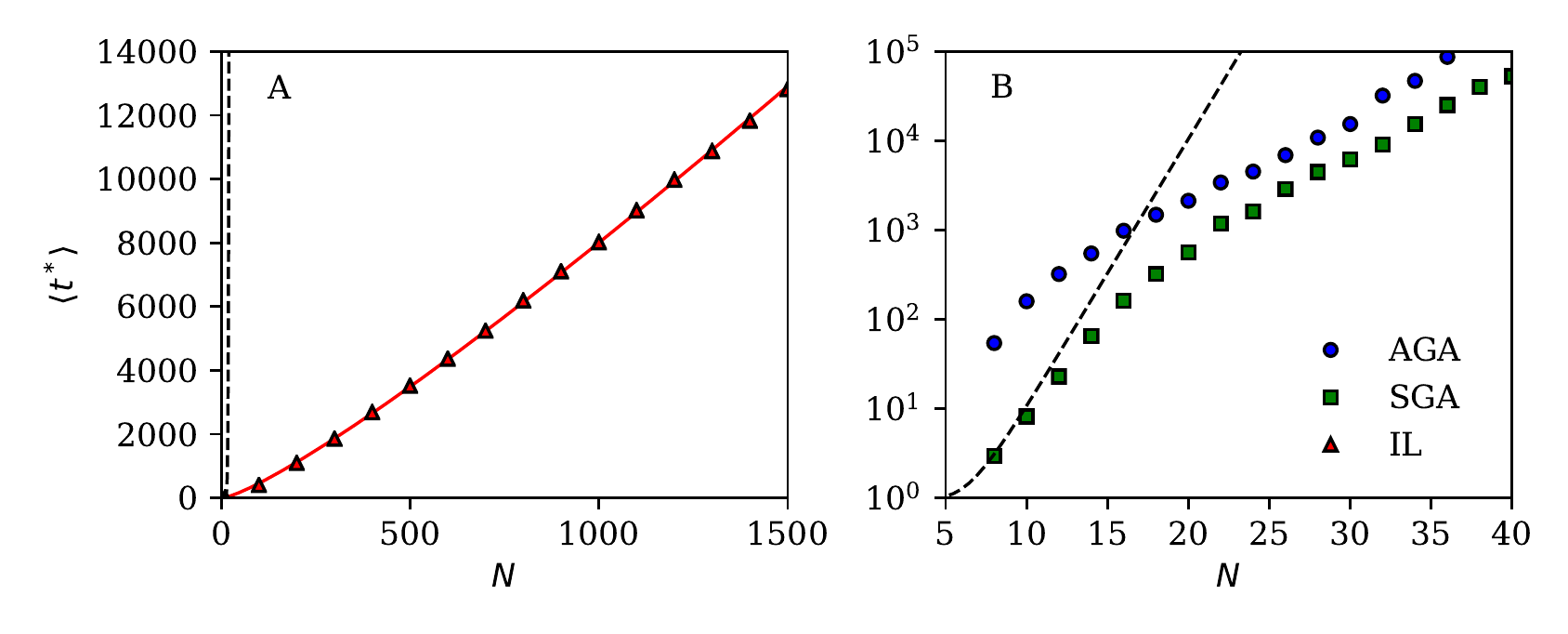}
\caption{Scaling of the mean halting time   $\langle t^* \rangle$  with the (smooth) landscape dimensionality $N$ for the  imitative learning (IL) search (panel A) and for  the asexual (AGA)  and  the  sexual  (SGA) genetic algorithms (panel B) as indicated.  The solid curve fitting the results of the IL search  is  the function $a N + b N \ln N$ with $ a = -2.30$ and $b=1.49$. 
The dashed curves  are the analytical prediction for the blind search (BS),  equation  (\ref{eq:t}).  The mean number of mutations per string  is $uN=0.01$ and  the population size is $M=100$.     }
\label{fig3}
\end{figure*}

In order to better quantify the effect of the  landscape dimensionality on the performance of the algorithms, Figure \ref{fig3} shows how the  mean halting time $\langle t^* \rangle$   scales  with $N$.  For the IL search  we find that $\langle t^* \rangle \asymp N \ln N $ similarly to the findings for the
random  adaptive   walk \cite{Fontanari_91,Flyvbjerg_92}, whereas for the evolutionary algorithms our results are inconclusive due to the 
limited range of values of  $N$ considered. However, our guess is that  $\langle t^* \rangle$ increases exponentially with increasing $N$  for those algorithms as we will argue below.  

The main reason for the superior performance of the IL over the evolutionary algorithms in finding the single maximum of  NK landscapes with $K=0$, especially for   high-dimensional landscapes, is that the genetic roulette is very  inefficient  to pick up the fittest string in a situation where the strings exhibit  similar fitness values. In fact, for  large $N$, the fitness difference between states that differ by a few bits is vanishingly small so the evolutionary algorithms are essentially exploring a nearly flat landscape, hence our conjecture that their mean halting times increase exponentially with the landscape dimensionality.  A similar result appears in the context of probabilistic adaptive walks, where it has been found  that the selection strength must grow logarithmically with $N$  in order that  the walker reaches the fitness maximum efficiently. Otherwise, the walker  cannot efficiently find the maximum, as the time required to reach it  becomes exponential in $N$ with overwhelming probability \cite{Heredia_17}. We stress, however, that  the evolutionary algorithms perform similarly to or poorer than the  blind search  for small  $N$ only.  For large $N$, they greatly outperform the blind search even though $\langle t^* \rangle$ seems to grow exponentially with $N$ (see Figure \ref{fig3}). In the Appendix we show that these conclusions  hold true for the Ising model of ferromagnetism as well. In particular,  for the non-interacting version, where the landscape exhibits  a single maximum, we find the same results as those described here for  $K=0$. For the ferromagnetic version, where the landscape exhibits  two degenerate  maxima, we find that all the cooperative search algorithms considered seem to be exponential in $N$.

 In contrast to the evolutionary algorithms,  the imitative learning  search always selects the fittest string as the model string, since its selection criterion uses  the  fitness rank rather than the relative fitness.  This is, of course,  a huge leverage for smooth landscapes, as shown in  Figures \ref{fig2} and  \ref{fig3}, because the  fitness  value is a reliable indicator of  proximity to the global maximum in this  case.
Next we will see whether this leverage holds for rugged landscapes as well. 

\subsection{Rugged landscapes}

\begin{figure*}[t!]
\centering  
\includegraphics[width=0.85\textwidth]{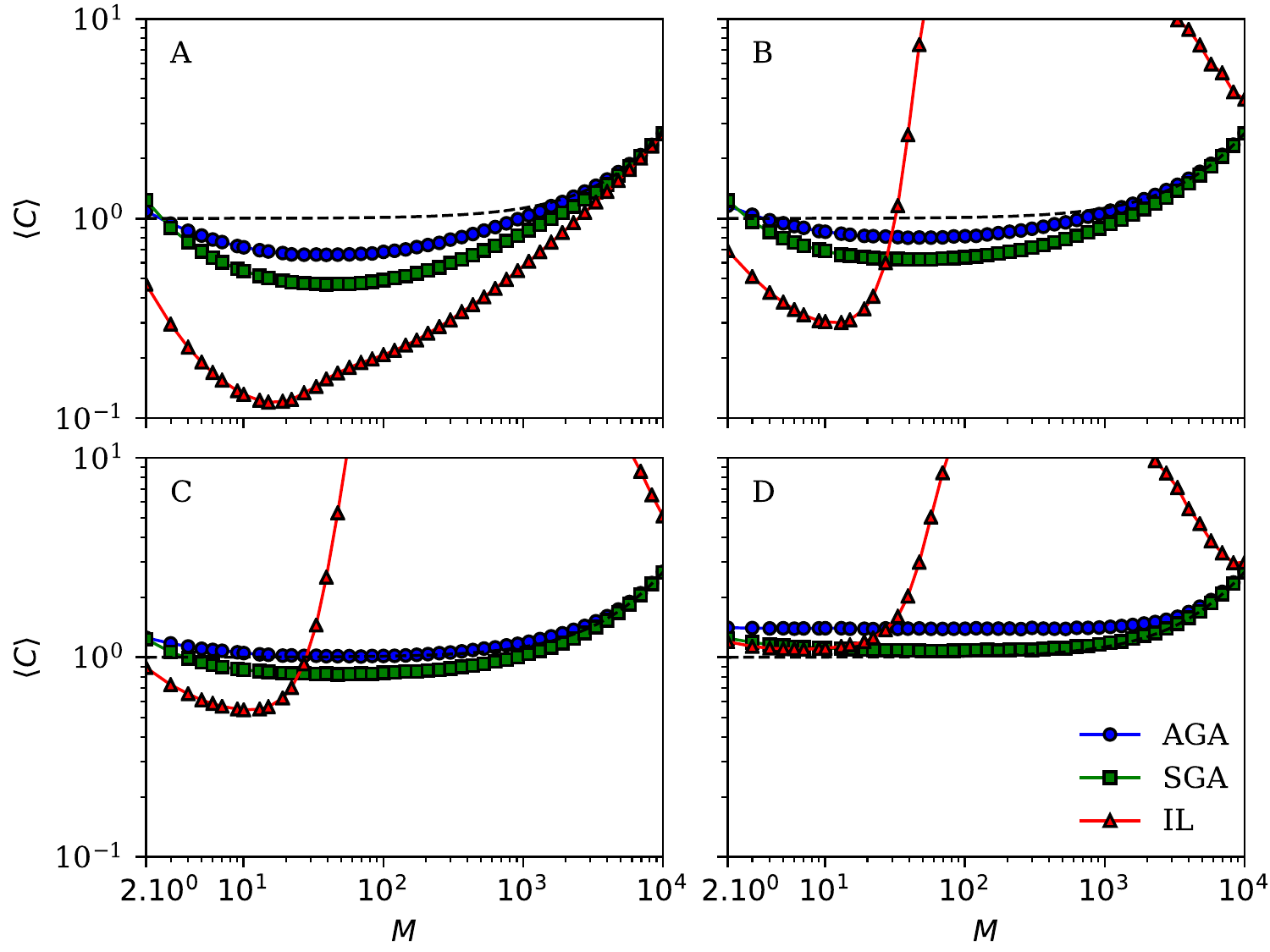}
\caption{Mean computational cost $\langle C \rangle $ as function of the  population size $M$  for rugged landscapes with fixed dimensionality $N=12$ and epistasis parameter $K=1$ (panel A), $K=3$ (panel B), $K=5$ (panel C) and $K=9$ (panel D).
  The dashed curve is the analytical prediction for the blind search, equation (\ref{Cind}), and the symbols are the simulation results for the imitative learning (IL) search, the asexual (AGA) and the  sexual (SGA) genetic algorithms as indicated.
The bitwise probability of mutation is $u=0.1$.  }
\label{fig4}
\end{figure*}

Figure \ref{fig4} exhibits the performances of the  search algorithms for landscapes of fixed dimensionality $N=12$ and increasing ruggedness, as determined by
the increasing values of the epistasis parameter $K=1,3,5,9$. These results reveal a  distinctive characteristic of the IL  search on rugged landscapes, namely, the appearance of a peak on the  computational cost for  large population 
 sizes \cite{Fontanari_14,Fontanari_15}.  In contrast,  increase of the  ruggedness of the fitness landscape 
does not produce qualitative changes on the dependence of the computational cost on the parameters $M$ and $u$ for  the evolutionary algorithms.

The poor performance of the IL search for large population sizes   is akin to the groupthink  phenomenon  of social psychology \cite{Janis_82}, which  occur  when everyone in a group starts  thinking alike as the result of people putting unlimited faith in a  leader (the model agent in our scenario).
In the IL search,   this phenomenon is  due to the rapid loss of diversity of the population  that occurs  when the model string is a  high fitness local maximum and the imitation process starts to produce too many clones of that string. (This explains why this effect does not appear for small $M$.)  This extreme  susceptibility  to the presence of local maxima is the price that the IL  search pays  for its good performance on smooth landscapes.  The groupthink-like phenomenon  can be circumvented by limiting the influence of the model agent using, for instance,  low connectivity networks \cite{Rodrigues_16} or  by allowing the agents to move randomly in an two-dimensional  space \cite{Gomes_19}. 
It is curious to note that a similar performance degradation  was reported in models of Parkinson's law that show the lessening of bureaucratic efficiency when  the size of administrative staff exceeds a certain number \cite{Klimek_09}.

As expected, the performances of the cooperative search algorithms degrade gradually as  the landscapes become more  rugged. In fact, since the state space size is fixed in  Figure \ref{fig4},   the density  of local maxima increases with increasing  $K$.   It is interesting that for the  rugged landscapes with $N=12$ and $K=9$ (panel D of Figure \ref{fig4}), for which the fitness correlation  between neighboring strings is $1/6$, the blind search equals or outperforms the cooperative search algorithms for all population sizes.

\begin{figure*}[t!]
\centering  
\includegraphics[width=0.85\textwidth]{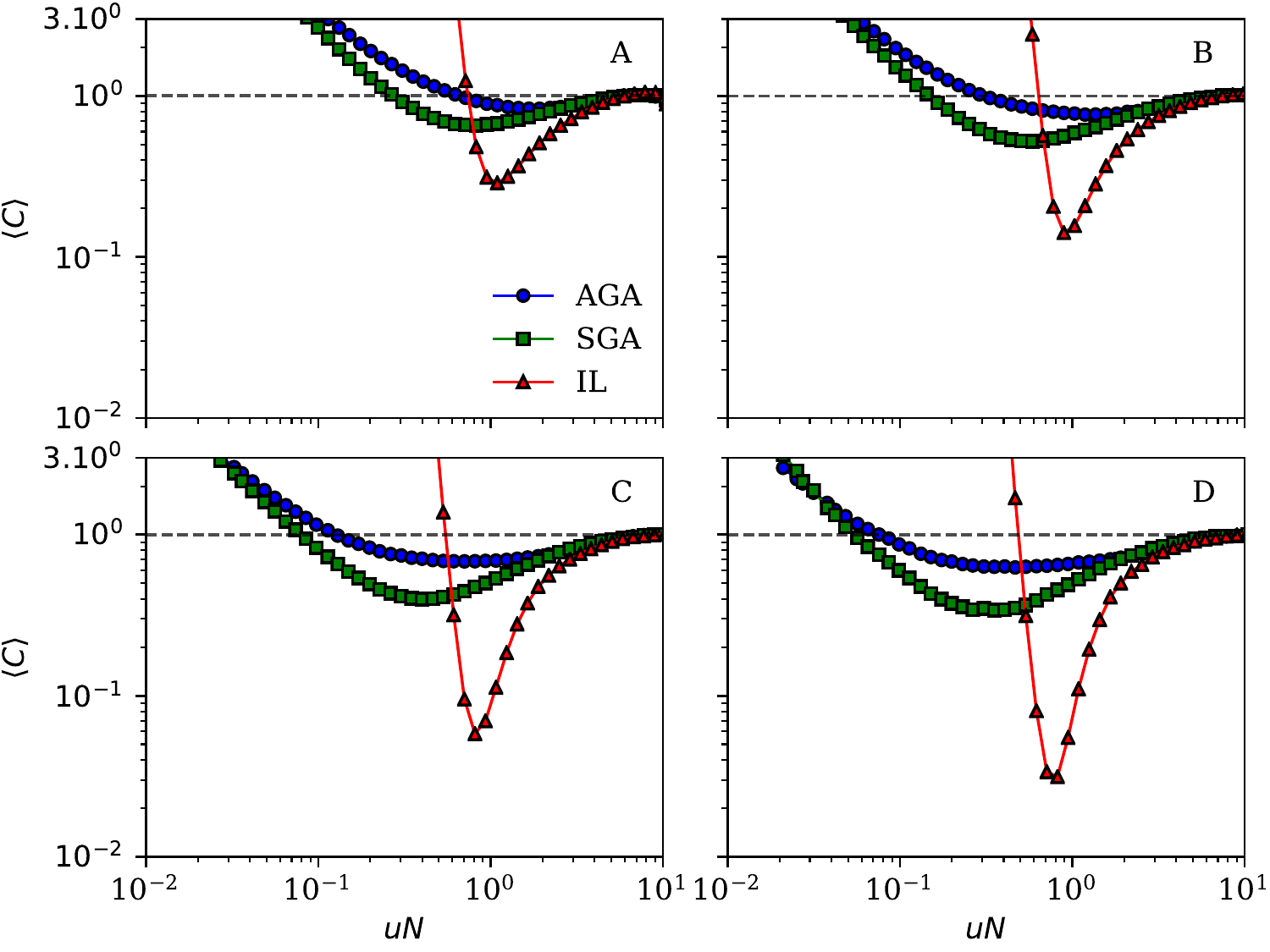}
\caption{Mean computational cost $\langle C \rangle$ as function of the mean number of mutations per string $u N$ for  rugged landscapes with parameters $N=12, K=3$ (panel A),  $N=15, K=4$ (panel B), $N=18, K=5$ (panel C) and   $N=21, K=6$ (panel D). 
 The  population size is $M=10$. The dashed straight lines indicate the performance of the blind search and the symbols are the simulation results for the imitative learning (IL) search, the asexual (AGA) and the sexual (SGA) genetic algorithms as indicated. }
\label{fig5}
\end{figure*}

Figure \ref{fig5} shows the performances of the search algorithms when the parameters $N$  and $K$  increase such that the  correlation between the fitness of neighboring strings, equation (\ref{eq:corr}), is kept fixed to $2/3$. The results lay bare the reliance of the cooperative search algorithms on the mutation operation  to produce diversity at the bit level and that the IL search is much more susceptible to lose diversity at that level and  to get trapped in the local maxima. We recall that if all strings exhibit the same bit at a given position then, in the absence of mutations, this bit will be fixed in the population since neither the crossover nor the imitation operation can flip it. This is a major hurdle if the fixed bit is not the correct one, i.e,  the corresponding bit in the global maximum, as indicated by the divergence of the computational cost as $u \to 0$ for all three cooperative algorithms.  

Another interesting result  exhibited in Figure \ref{fig5} is that the task seems to become easier as the dimensionality of the landscape increases. This is so because the density of local maxima actually decreases as both $N$ and $K$ increase such that the ratio $(K+1)/N$  is kept constant. For example, for the ensemble of 100 landscapes with $N=12$ and $K=3$ we find that the mean number of maxima is $27.18$ and the density of maxima is $0.007$, whereas for the landscapes with  $N=21$ and $K=6$ the mean  number of maxima is $3074.52$, which corresponds to a density  of $0.001$ maxima per state. In fact, in the limits of  large $K$ and  $N$ such that the ratio $\alpha = K/N$ is fixed,   the probability that a random string is a local maximum can be analytically shown to decay algebraically as $N^{-1/\alpha}$ \cite{Hwang_18}, so that the density of local maxima must vanish for large $N$.
This means that, as $N$  and $K$ increase, the number of paths (i.e., learning trajectories)  leading to the global maximum that circumvent  the local maxima increases very fast \cite{Campos_19}, explaining thus the qualitative similarity of the performances of the algorithms  illustrated  in Figure \ref{fig5}  for rugged landscapes to those  in Figure \ref{fig2} for smooth landscapes. In this line, it is interesting to note that the local maxima of the NK-fitness landscapes are strongly clustered in the state space  \cite{Nowak_15}.

One may argue  that since in real-world problems the state space is very large, the odds  that a general-purpose search algorithm reaches the optimal solution  in all runs  within a feasible runtime are  negligible for NP-com\-plete  problems. Hence, since it is impractical to estimate the mean halting time $\langle t^* \rangle$ for problems with a very large state space, the computational costs exhibited in, say,  Figure \ref{fig4} may not be indicative of the performance of the cooperative search algorithms in the real-world scenario where the duration of the search is limited a priori. 
We can  partially assess the significance of this claim  by estimating the fraction of runs for which  the global maximum is found  when the  duration of the search is fixed to $t$. In the case of the blind search, this fraction is nothing but the probability $\pi_M (t)$ that the population of $M$ agents finds the global maximum before or at time $t$ given in  equation  (\ref{eq:PM}).

\begin{figure*}[t!]
\centering  
\includegraphics[width=0.88\textwidth]{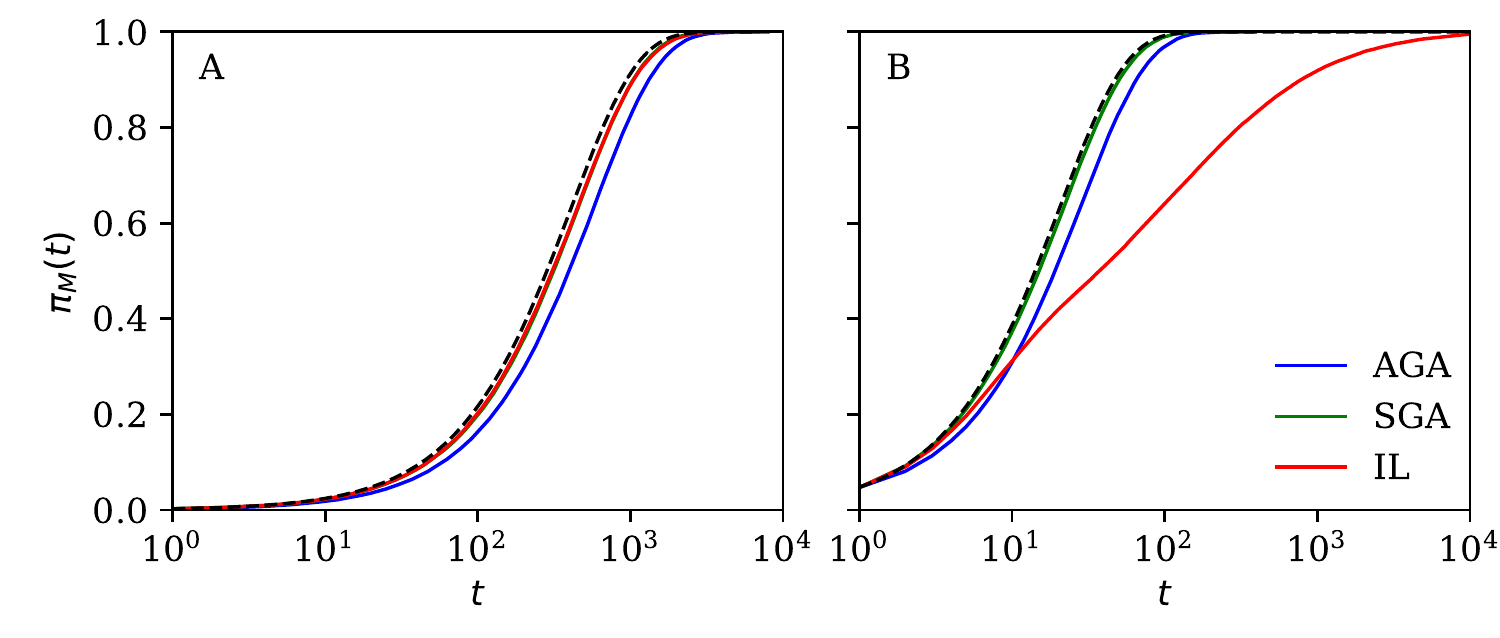}
\caption{Fraction of  runs  for  which the global maximum is found before or at time $t$  for the imitative learning  (IL) search, the asexual (AGA) and the  sexual (SGA) genetic algorithms as indicated. The dashed curve is the analytical prediction for the blind search, equation  (\ref{eq:PiM}). The bitwise mutation probability is $u=0.1$ and  the population size is $M=10$ (panel A)   and  $M=200$  (panel B).  The curves for the IL search and SGA are indistinguishable in panel A.  The parameters of the NK-fitness landscapes are  $N=12, K=9$ so that the mean number of mutations per string is  $u N = 1.2$.   }
\label{fig6}
\end{figure*}

Accordingly, Figure \ref{fig6} shows  $\pi_M (t)$  for  the challenging case exhibited in panel D of  Figure \ref{fig4} where the blind search either matches or outperforms all three cooperative search  algorithms.  The results indicate that this outcome holds true for searches of limited duration as well.  We note that, similarly to the  observed in the study of  searches of unrestricted duration,  the short-time performance of the IL search  is strongly dependent on the population size whereas the evolutionary algorithms are little influenced by this parameter.

\begin{figure*}[t!]
\centering  
\includegraphics[width=0.88\textwidth]{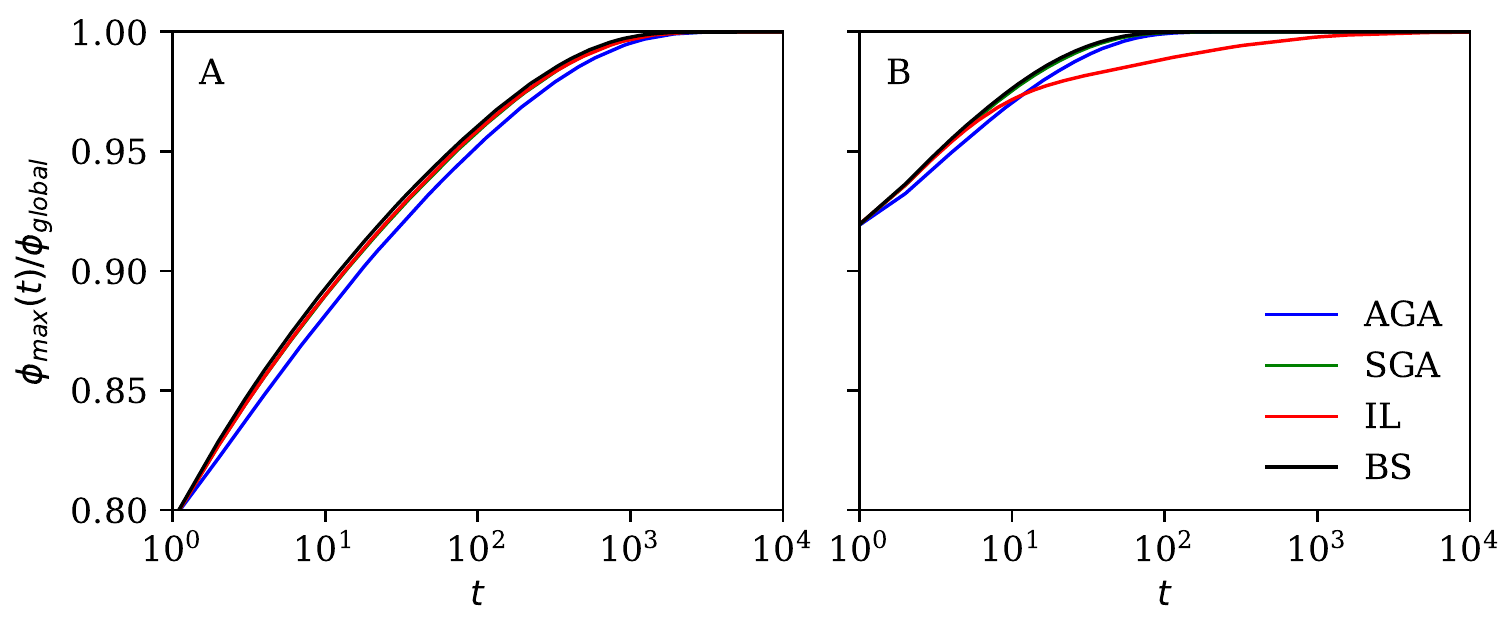}
\caption{Ratio between the fittest state found up to time $t$, $\Phi_{\text{max}}$, and the global optimum fitness,
$\Phi_{\text{global}}$, for the imitative learning (IL) search, the blind search (BS), the  asexual (AGA)  and the sexual (SGA) genetic algorithms as indicated.   The bitwise mutation probability is $u=0.1$ and  the population size is $M=10$  (panel A) and  $M=200$  (panel B). The curves for the IL search and SGA are indistinguishable in panel A, whereas the curves for the BS and SGA are indistinguishable in panel B. The parameters of the NK-fitness landscapes are $N=12, K=9$ so that the mean number of mutations per string is  $u N = 1.2$.   }
\label{fig7}
\end{figure*}

In a similar vein, an argument  that   is usually put forth in support of general-purpose optimization algorithms is that they excel at finding solutions that are `sufficiently good' for practical purposes within realistic  runtimes. In this line of reasoning, finding  optimal or near-optimal solutions were never the intent of the bio and cultural inspired algorithms.  This is a fair point,  though we think it is  instructive, to say the least,  to learn how those algorithms fare when their task is to find the optimal solution, as done in this paper.  However, we can easily check the validity of this argument  by measuring  the expected value of the fittest solution found by the algorithms up to time $t$, which we denote by 
$\Phi_{\text{max}}(t)$. More pointedly,  for a fixed landscape we run $10^3$ searches and  record the fittest solution found up to  time $t$ for each run. Then we average over these fitness values to obtain $\Phi_{\text{max}}(t)$ for a fixed landscape.
Since we also need to average this result over different landscapes, we consider a more appropriate,  landscape-independent measure    given by  the ratio $\Phi_{\text{max}}/\Phi_{\text{global}} \leq 1$, where $\Phi_{\text{global}}$ is the fitness of the global maximum.

Accordingly, Figure \ref{fig7} shows the ratio $\Phi_{\text{max}}/\Phi_{\text{global}}$  for the challenging rugged landscapes of panel D of Figure \ref{fig4}. 
We note that, in this case, the results for the blind search were obtained by simulations as the analytical results of Subsection \ref{BS} deal with  the distributions of the absorbing times only. These illuminating results  show no evidence that the cooperative algorithms are more effective than the blind search in finding good solutions in searches of  short  duration. In fact, the very same conclusions hold true when the task changes from finding the global maximum in the minimum runtime to finding high fitness states within a fixed runtime.  Since for large $M$ there is a good chance of finding high fitness states already during  the assemblage of the initial population, we have  $\Phi_{\text{max}}(0)/\Phi_{\text{global}} \to 1$  as $M$ increases, which explains the great difference in the ranges of  variation  of the ratio $\Phi_{\text{max}}/\Phi_{\text{global}}$  between the two panels of  Figure \ref{fig7}. Therefore our results support the (actually) conservative stance  that the algorithms that  reach the global maxima more rapidly are also more likely to visit the fittest states within a fixed runtime.

\section{Conclusion}\label{sec:Conc}

The imitative learning (IL) search was introduced originally as a model to study quantitatively the potential of imitation as the underlying mechanism -- the critical connector -- of collective brains \cite{Fontanari_14,Fontanari_15}.  A natural baseline to assess that potential is a scenario where the agents explore the problem space independently of each other, performing a sort of blind search on that space. The finding that IL  search performs much worse than the baseline for certain values of the control parameters  and  the apparent similarities  between the IL and the popular evolutionary algorithms gave rise to  the question  of whether similar negative results could hold for those algorithms as well.
 The aim of this paper is to  address this issue   by challenging  the cultural and biologically inspired cooperative search algorithms to beat the blind search in the task of  finding the unique global maximum of NK-fitness landscapes \cite{Kauffman_89}. 
 
 In addition to the IL  search, we consider two evolutionary algorithms, viz., the asexual (AGA) and the sexual  (SGA) genetic algorithms. The difference between them is that the former lacks the crossover mechanism to generate diversity in the string population.
 The main performance measure considered here is the total number of agent updates required by the algorithms  to find the global maximum of the fitness landscape. Our conclusion  is that  the cooperative search algorithms are only marginally superior to the blind search  in the exploration of  rugged NK-fitness landscapes. Moreover, the evolutionary algorithms do not exhibit the catastrophic performance of the  imitative learning  search that is observed for certain population sizes  and that has been likened to the  groupthink phenomenon of social psychology \cite{Janis_82}.

 Within a genetic perspective,   the model string in the IL search may be thought of as a mandatory parent in all mates  at a given generation, which contributes  a single bit to the offspring. In the absence of mutations, this offspring is identical to the other parent,  namely, the randomly chosen target string, except for the bit that comes from the model string. In addition, that bit is not random  since it must differ from the original bit of  the target string. In fact,  since the imitation process was based on Axelrod's model of cultural dissemination \cite{Axelrod_97},  the IL   search follows the rules of cultural, rather than of genetic, evolution.

Our definition of computational cost   as, essentially, the total number of updates performed by the algorithms to find the optimal solution    begets a humbler perspective on  the power of general-purpose search algorithms (see \cite{Wolpert_97} for another approach to this issue that leads to a similar conclusion). In particular, we find that even for simple problems with no local maxima (i.e., $K=0$), the evolutionary algorithms are  not   much better than the blind search provided the landscape dimensionality is not too large.  In this simple scenario, the IL exhibits the
best performance since it always guarantees the  selection of the fittest string as a model to be imitated, thus avoiding the  genetic drift that hinders the  performance of the evolutionary algorithms  (see Figures  \ref{fig2} and \ref{fig3}). The robustness of these findings are confirmed through the analysis of the Ising model of ferromagnetism offered in the Appendix.

The prospects of the cooperative search algorithms  are somewhat gloomy in the case of  rugged landscapes plagued with local maxima that may act as traps for the evolving population.  In fact, even for mildly rugged landscapes (see panel D of Figure \ref{fig4}) the blind search either outperforms or matches the cooperative search algorithms regardless of whether the duration of the search for the global maximum is limited a priori  or not.  This conclusion holds true also in the case the criterion to evaluate the algorithms is the fittest state found within a fixed runtime.

In many respects, the unenthusiastic performances of the cooperative search algorithms reported here, as compared with the baseline set by the blind search,   are  in line with the very notion of NP-completeness class \cite{Garey_79}  in the sense that we should not expect any algorithm to change substantially  the scale of the time needed to find the optimal solution for problems in that class. (Here we assume  without proof that the adjacent neighborhood variant of the NK-model with $\alpha = K/N > 0$ fixed  is NP-complete.)
 Overall we find that if the population size $M$ and the bitwise mutation probability $u$ are  tuned so as to optimize the performance of each cooperative search algorithm separately,  then the IL  surpasses the evolutionary algorithms, especially for rugged landscapes  characterized by a low density of local maxima as illustrated in  Figure \ref{fig5}.

The somewhat unexpected conclusions of our study  calls into question the efficiency of  the mechanisms of selection and recombination to explore NK-fitness landscapes. This is probably due to a peculiarity of these landscapes, viz.,  they become flatter as their dimensionality increase, thus greatly   impairing the capacity of the genetic roulette to select the best  agents for the next generation. 
 We stress, however, that the performance advantage of the evolutionary algorithms  over the blind search actually increases with the landscape dimensionality (see Figure \ref{fig3}). As shown in the Appendix, the same  conclusions hold true for a far more popular fitness landscape, the Ising model of ferromagnetism. We see the poor performance of the evolutionary algorithms on single-peak fitness landscapes   as the price   general-purpose algorithms have to pay  to deliver good solutions using hardly any  information on the optimization problems they are set to tackle. Like the blind search, for those algorithms the scaling on the landscape dimensionality of the time required to find the global maxima is not very sensitive to the topology of  the landscape. This contrast with the imitative learning search that performs almost optimally  for single-peak landscapes but exhibits a  catastrophic performance for landscapes with local maxima if its parameters are not  tuned    properly. There is indeed no free lunch in optimization.

\bigskip

\acknowledgments
The research of JFF was  supported in part 
 by Grant No.\  2017/23288-0, Fun\-da\-\c{c}\~ao de Amparo \`a Pesquisa do Estado de S\~ao Paulo 
(FAPESP) and  by Grant No.\ 305058/2017-7, Conselho Nacional de Desenvolvimento 
Cient\'{\i}\-fi\-co e Tecnol\'ogico (CNPq).
SMR  was supported by grant  	15/17277-0, Fun\-da\-\c{c}\~ao de Amparo \`a Pesquisa do Estado de S\~ao Paulo 
(FAPESP)  and by the Coordena\c{c}\~ao de Aperfei\c{c}oamento de Pessoal de N\'{\i}vel Superior - Brasil (CAPES) - Finance Code 001. LFA was supported by a CNPq fellowship.

\section*{Appendix}

\renewcommand{\thefigure}{A\arabic{figure}}
\renewcommand{\theequation}{A\arabic{equation}}
\setcounter{figure}{0}
\setcounter{equation}{0}

In this Appendix we address the robustness of our findings about the performance of the cooperative search algorithms on smooth fitness landscapes.  Although the NK-model of rugged landscapes is widely used in a variety of disparate disciplines, such as evolutionary biology, physics and economics, the wanting performance of the  evolutionary algorithms reported in this paper  may raise concern about the usefulness of  the NK  model to describe realistic landscapes. To dispel these doubts, here we briefly examine   
the performance of the cooperative  search algorithms on  a  very popular fitness (or energy) landscape, viz.,  the Ising  model of ferromagnetism \cite{Huang_63}.

\begin{figure}
\centering  
\includegraphics[width=0.47\textwidth]{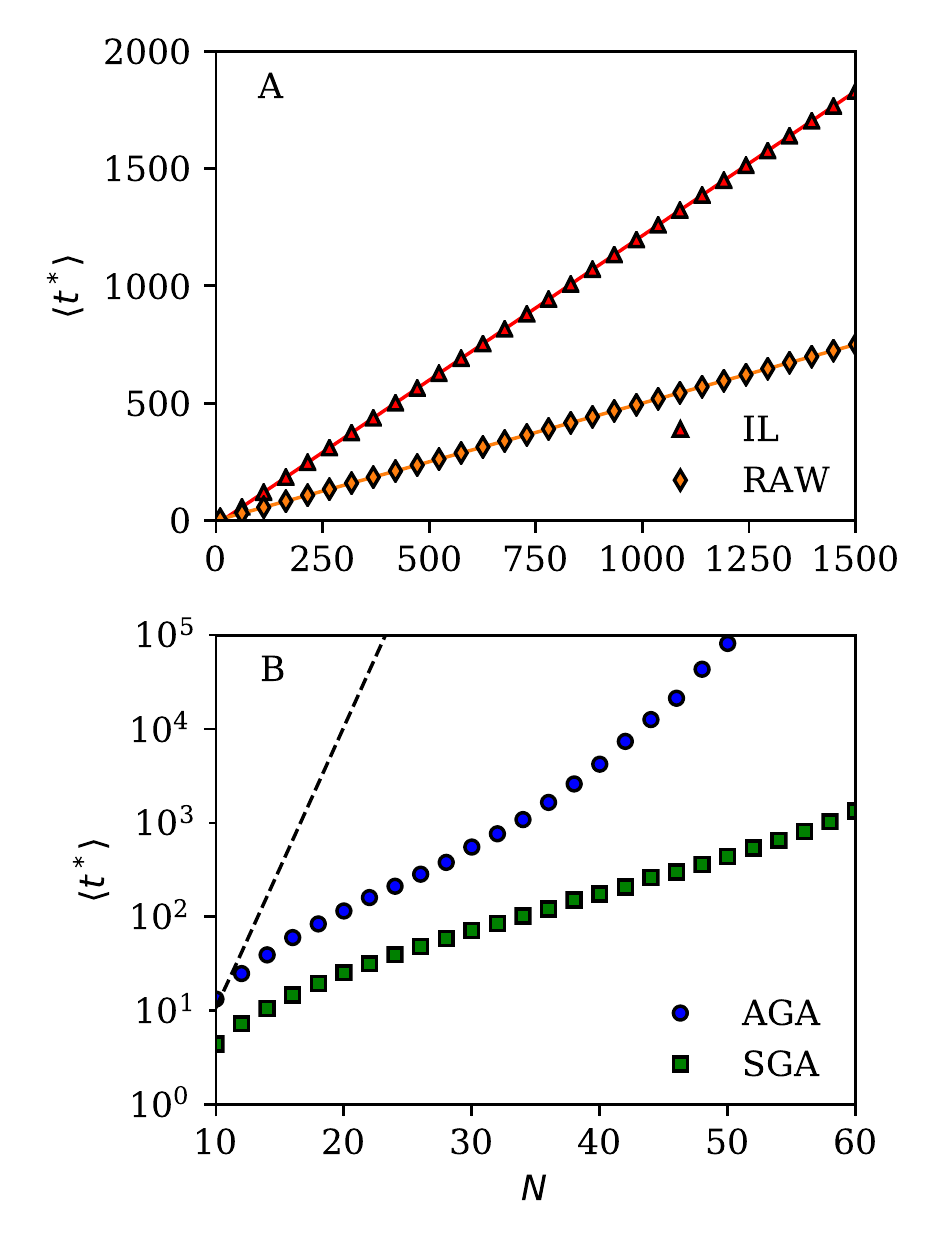}
\caption{Scaling of the mean halting time   $\langle t^* \rangle$  on the  landscape dimensionality $N$
for the  non-interacting Ising model.  The  results for the imitative learning (IL) search and the random adaptive  walk (RAW) are shown in  panel A, whereas  the results for the  asexual (AGA)  and  the  sexual  (SGA) genetic algorithms are shown in panel B, as indicated.  The solid curve fitting the result of the IL search  is  the function $a N +b$ with $ a = 1.23$  and $b=-18.13$, whereas the solid curve fitting the result of the RAW is $N/2$.
The dashed curve  is the analytical prediction for the blind search (BS),  equation  (\ref{eq:t}).  The mean number of mutations per string  is $uN=0.1$ and  the population size is $M=100$.  }
\label{figA1}
\end{figure}

We begin with the non-interaction version of the Ising model where the fitness (the opposite of the energy) of the system  specified by the binary string $\mathbf{x} = \left ( x_1, x_2, \ldots,x_N \right )$ with
$x_i = 0,1$ is 
\begin{equation}\label{ni}
F_{ni} = \sum_{i=1}^N \left ( 2 x_i - 1\right)  + N + 1 .
\end{equation}
Here the spin variable is $s_i = \left ( 2 x_i - 1\right) = \pm 1$ and 
we have added the factor $N + 1$ into the fitness definition  so as to guarantee that it is positive for all configurations and hence that the genetic roulette behind the evolutionary algorithms can be applied straightaway. Clearly, this fitness function has a single maximum at $x_i = 1 ~\forall i$  and so it offers an alternative to the NK-fitness landscape with $K=0$.

Figure \ref{figA1} shows how the  mean halting time $\langle t^* \rangle$   scales  with the  landscape dimensionality $N$ for the fitness function defined in equation (\ref{ni}).  In addition to the  four search algorithms discussed in the paper, this figure shows the performance of the random adaptive   walk (RAW) as well \cite{Kauffman_87,Nowak_15}. This search algorithm considers a single agent (walker) that at each time step  flips a randomly chosen  bit  that  does not 
result in a fitness decrease. For the fitness function (\ref{ni}),  this strategy is equivalent to the greedy heuristics, since all bit flips that do not decrease the fitness increase it by the same amount. Hence we have  
 $\langle t^* \rangle = N/2$ for the RAW.   Interestingly, we find that the mean halting  time of the IL  search also scales linearly with  $N$, i.e.,  $\langle t^* \rangle \asymp N  $.   Similarly to the findings summarized in  Figure \ref{fig3} for the case $K=0$, the results for  the evolutionary algorithms are inconclusive due to the 
limited range of values of  $N$ considered. Nonetheless,  it is clear that the performances of those algorithms scale very poorly with the  landscape dimensionality. Therefore, the conclusions drawn from the analysis of smooth NK-fitness landscapes hold true for the well-known case  of the non-interacting Ising model landscape.
 
\begin{figure}
\centering  
\includegraphics[width=0.47\textwidth]{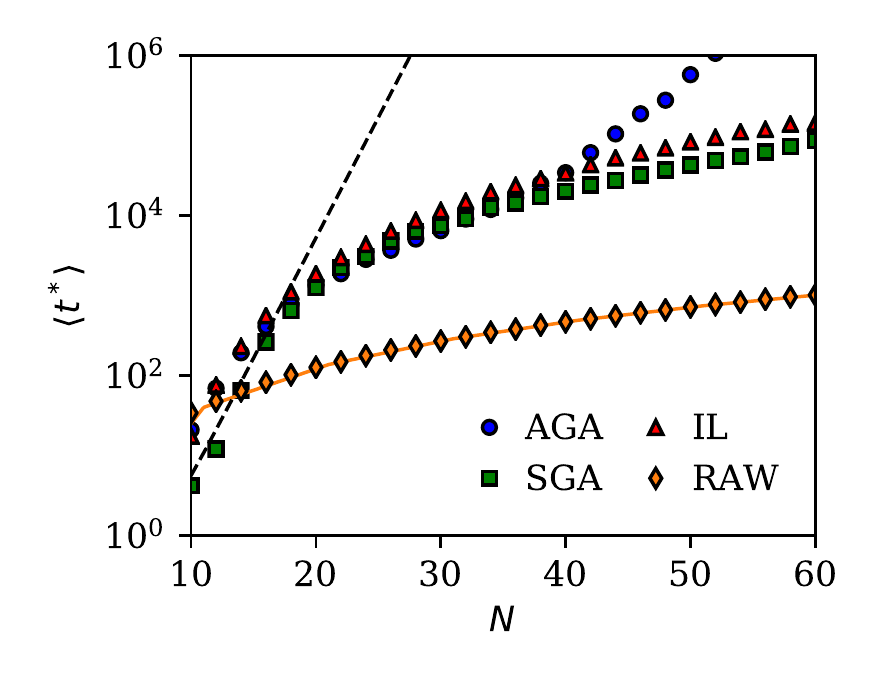}
\caption{Scaling of the mean halting time   $\langle t^* \rangle$  on the  landscape dimensionality $N$
for the  one-dimensional ferromagnetic  Ising model.   The solid curve fitting the result of the RAW  is  the function $0.41 N^2$ whereas 
the dashed curve  is the analytical prediction for the blind search (BS),  equation  (\ref{eq:t}), with $p=1/2^{N-1}$.  The mean number of mutations per string  is $uN=0.1$ and  the population size is $M=100$.  }
\label{figA2}
\end{figure}
 
 We consider now the one-dimensional ferromagnetic Ising model where the fitness of the system  specified by the binary string $\mathbf{x}$ is
\begin{equation}\label{f}
F_{f} = \sum_{i=1}^N \left ( 2 x_i - 1\right) \left ( 2 x_{i+1} - 1\right)  + N + 1 
\end{equation}
with $x_{N+1} \equiv x_1$. This fitness function has two degenerate maxima, viz.,  $x_i=1 ~ \forall i$ and $x_i=-1 ~ \forall i$. Moreover, it has several plateaus where a given configuration and its $N$ neighbors exhibit the same fitness. Actually, the reason we tolerate flips that do not increase fitness in the  RAW  is to allow the walker to escape those plateaus.  The results summarized in Figure \ref{figA2} show the disruptive effect of the two degenerate maxima on the performance of all the cooperative search algorithms considered. Since the fitness function (\ref{f}) has no local maxima, the reason for the poor performance of the IL search  is probably the frequent alternation of the model agent between  strings in the  neighborhoods of the two degenerate maxima. The back and forth motion between those regions, as well as the blind search on the fitness plateaus,  are  the reasons that the halting time of the RAW scales with $N^2$, rather than with $N$,  for this landscape. 

Finally, we note that the  reason the RAW performs better than the cooperative search algorithms on the two landscapes considered in this Appendix is due solely to the absence of local maxima on those landscapes.  We stress that  despite the somewhat wanting performances of the cooperative search algorithms regarding their scaling with the landscape dimensionality, they  do much better than the blind search in the case of  high dimensional landscapes.

\end{document}